\documentclass{article}

\PassOptionsToPackage{numbers, sort&compress}{natbib}

\usepackage[final]{neurips_2022}

\usepackage[utf8]{inputenc} % allow utf-8 input
\usepackage[T1]{fontenc}    % use 8-bit T1 fonts
\usepackage{hyperref}       % hyperlinks
\usepackage{url}            % simple URL typesetting
\usepackage{booktabs}       % professional-quality tables
\usepackage{amsfonts}       % blackboard math symbols
\usepackage{nicefrac}       % compact symbols for 1/2, etc.
\usepackage{microtype}      % microtypography
\usepackage{xcolor}         % colors
\usepackage{multirow}
\usepackage{graphicx}

\title{COVIDx~CXR-3: A Large-Scale, Open-Source Benchmark Dataset of Chest X-ray Images for Computer-Aided COVID-19 Diagnostics}

\author{
    Maya Pavlova$^{1,2}$, Tia Tuinstra$^{1,2}$, Hossein Aboutalebi$^{1,3,4}$,\\
    \textbf{Andy Zhao$^{1,2}$, Hayden Gunraj$^{1,2}$, Alexander Wong$^{1,2,4,5}$}\\
    $^{1}$Vision and Image Processing Lab, University of Waterloo\\
    $^{2}$Department of Systems Design Engineering, University of Waterloo\\
    $^{3}$Cheriton School of Computer Science, University of Waterloo\\
    $^{4}$Waterloo AI Institute, University of Waterloo\\
    $^{5}$DarwinAI Corp.\\
    \texttt{\{mspavlova, ttuinstra, haboutal, andy.zhao,}\\
    \texttt{hayden.gunraj, a28wong\}@uwaterloo.ca}
}

\begin{document}

\maketitle

\begin{abstract}
    After more than two years since the beginning of the COVID-19 pandemic, the pressure of this crisis continues to devastate globally. The use of chest X-ray (CXR) imaging as a complementary screening strategy to RT-PCR testing is not only prevailing but has greatly increased due to its routine clinical use for respiratory complaints. Thus far, many visual perception models have been proposed for COVID-19 screening based on CXR imaging. Nevertheless, the accuracy and the generalization capacity of these models are very much dependent on the diversity and the size of the dataset they were trained on. Motivated by this, we introduce COVIDx~CXR-3, a large-scale benchmark dataset of CXR images for supporting COVID-19 computer vision research. COVIDx~CXR-3 is composed of 30,386 CXR images from a multinational cohort of 17,026 patients from at least 51 countries, making it, to the best of our knowledge, the most extensive, most diverse COVID-19 CXR dataset in open access form. Here, we provide comprehensive details on the various aspects of the proposed dataset including patient demographics, imaging views, and infection types. The hope is that COVIDx~CXR-3 can assist scientists in advancing machine learning research against both the COVID-19 pandemic and related diseases.
\end{abstract}

\section{Introduction}
\vspace{-0.1in}
    As the COVID-19 pandemic continues to progress internationally, the push for effective and robust screening methods is imperative. One method that has seen significant increase in usage in the COVID-19 clinical workflow is chest X-ray (CXR) imaging, which offers complementary screening information to transcriptase-polymerase chain reaction (RT-PCR) testing with the benefits of more readily accessible and mobile healthcare equipment~\cite{Jacobi}. In addition, CXR imaging is routinely conducted in parallel to tedious viral testing for patients with respiratory complaint~\cite{BSTI} to reduce patient volume and provide more clinical detail. As such, multiple works have investigated deep learning-driven computer-aided diagnostics to both detect positive SARS-CoV-2 cases from CXR images~\cite{covidnet,medusa,cxr2} as well as provide severity assessments that can be leveraged in patient triaging~\cite{alex2020covidnets,aboutalebi2021covidnet,medusa}. Nevertheless, as a result of the rapid progression of the pandemic and drive for results, such studies have been limited in terms of quantity and/or diversity of patients as well as standardization between datasets for higher public accessibility. In this work, we introduce COVIDx~CXR-3, a large-scale benchmark dataset of CXR images that has been carefully curated with expert radiologists and clinicians for computer-aided COVID-19 research comprising of 30,386 CXR images from a multi-institutional, multinational cohort of 17,026 patients across at least 51 countries.  To the best of the authors' knowledge, COVIDx~CXR-3 is the most extensive, accessible COVID-19 CXR dataset in open-source form. Additionally, we examine the dataset diversity in regards to patient findings, demographics, and imaging view, and explore areas of potential bias and dataset imbalances.

\section{Methods}
\vspace{-0.1in}
    \begin{figure}[!t]
        \centering
        \vspace{-0.2in}
        \includegraphics[width=0.8\linewidth]{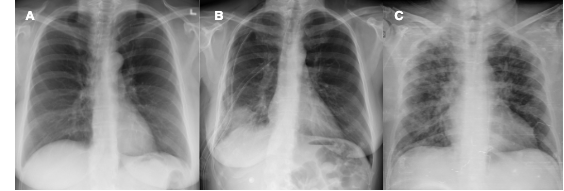}
        \caption{Example CXR images from COVIDx~CXR-3 benchmark dataset: (A) normal or no pneumonia or SARS-CoV-2 finding, (B) pneumonia infection, and (C) SARS-CoV-2 positive finding.}
        \label{fig:ex_cxr}
    \end{figure}

    The COVIDx~CXR-3 benchmark dataset comprises carefully curated CXR images from a multi-institutional, multinational patient cohort, in collaboration with expert radiologists and clinicians. More specifically, COVIDx~CXR-3 comprises patient cohorts from the following organizations and initiatives: (1) RSNA International COVID-19 Open Radiology Database (RICORD)~\cite{ricord}, (2) Stony Brook University (COVID-19-NY-SBU)~\cite{stonybrook}, (3) Valencian Region Medical Image Bank (BIMCV)~\cite{bimcv}, (4) RSNA Pneumonia Detection Challenge dataset~\cite{kaggle}, (5) COVID-19 Image Data Collection~\cite{Cohen}, (6) Fig.1 COVID-19 CXR Dataset Initiative~\cite{Figure1}, (7) ActualMed COVID-19 CXR Dataset Initiative~\cite{actualmed}, and (8) COVID-19 Radiography Database~\cite{kaggle2}. Expert feedback was used to evaluate the quality and applicability of the images from each source, leading to removal of some images from the listed sources. A test set that has no patient overlap with the training set was created from randomly sampled patients from the Radiological Society of North America (RSNA) RICORD~\cite{ricord} and Pneumonia Detection~\cite{kaggle} initiatives as a result of their high image and annotation quality. Each collected patient case is documented with one of the following findings: (A) normal or no pneumonia and no SARS-CoV-2 infection, (B) non-SARS-CoV-2 pneumonia infection, and (C) positive SARS-CoV-2 (COVID-19) infection. Example images from COVIDx~CXR-3 for each infection case are shown in Figure~\ref{fig:ex_cxr}.

\section{Results and Discussion}
\vspace{-0.1in}
    \begin{table}[t]
        \caption{Distribution of CXR images and patient cases (in parentheses) for both training and evaluation datasets in COVIDx~CXR-3, split by infection type. Note the total for patients is 10 less than the sum across the training set as some patients had both SARS-CoV-2 negative and positive CXR images from different studies.}
        \label{tab:splits}
        \centering
        \begin{tabular}{lcccc}
            \toprule
            & \multicolumn{3}{c}{\textbf{Infection Type}} & \\
            \cmidrule(lr){2-4}
            \textbf{Split} & Normal & Pneumonia & COVID-19 & \textbf{Total} \\
            \midrule
            Training & 8,437 (8,238) & 5,555 (5,612) & 15,994 (2,808) & 29,986  (16,648)\\
            \midrule
            Test & 100 (100) & 100 (100) & 200 (178) & 400 (378)\\
            \midrule
            \textbf{Total} & 8,537 (8,338) & 5655 (5,712) & 16,194 (2,986) & 30,386 (17,026)\\
            \bottomrule
        \end{tabular}
    \end{table}

\begin{table}[h] 
    \centering
    \caption{Summary of demographic and imaging protocol variables for the COVIDx~CXR-3 benchmark dataset. Age and sex statistics are expressed on a patient level, while imaging view statistics are expressed on an image level.}
    \label{tab:demographic}
    \begin{tabular}{|c|c||c|c|}
        \hline
        \multicolumn{2}{|l||}{\textbf{Age}} & \multicolumn{2}{l|}{\textbf{Sex}}\\ \hline
        <18 & 792 (4.7\%) & Male & 9,050 (53.2\%)\\
        {[18, 59]} & 11,275 (66.2\%) & Female & 6,830 (40.1\%)\\
        (59, 74] & 3,111 (18.3\%) & Unknown & 1,146 (6.7\%)\\
        \cline{3-4}
        (74, 90] & 702 (4.1\%) & \multicolumn{2}{l|}{\textbf{Imaging view}}\\
        \cline{3-4}
        >90 & 7 (0.04\%) & PA & 8,553 (28.1\%)\\
        Unknown & 1,139 (6.7\%) & AP & 20,372 (67.0\%)\\
        & & Unknown & 1,461 (4.8\%)\\ \hline
    \end{tabular}%
\end{table}

    Table~\ref{tab:splits} shows the distribution of CXR images and patients in regard to infection type. The final COVIDx~CXR-3 benchmark dataset is composed of 30,386 CXR images from 17,026 unique patients, of which 16,194 CXR images (53.3\%) and 2,986 patients (17.5\%) come from SARS-CoV-2 positive cases, creating a relatively balanced training and testing set for SARS-CoV-2 positive and negative detection in terms of image count. The smaller balanced test set is the result of sampling an 80\%/20\% patient split from the RSNA RICORD~\cite{ricord} initiative for positive SARS-CoV-2 cases to ensure networks are evaluated against expertly annotated positive samples. 

\begin{figure}[!t]
        \centering
        \includegraphics[width=0.5\linewidth]{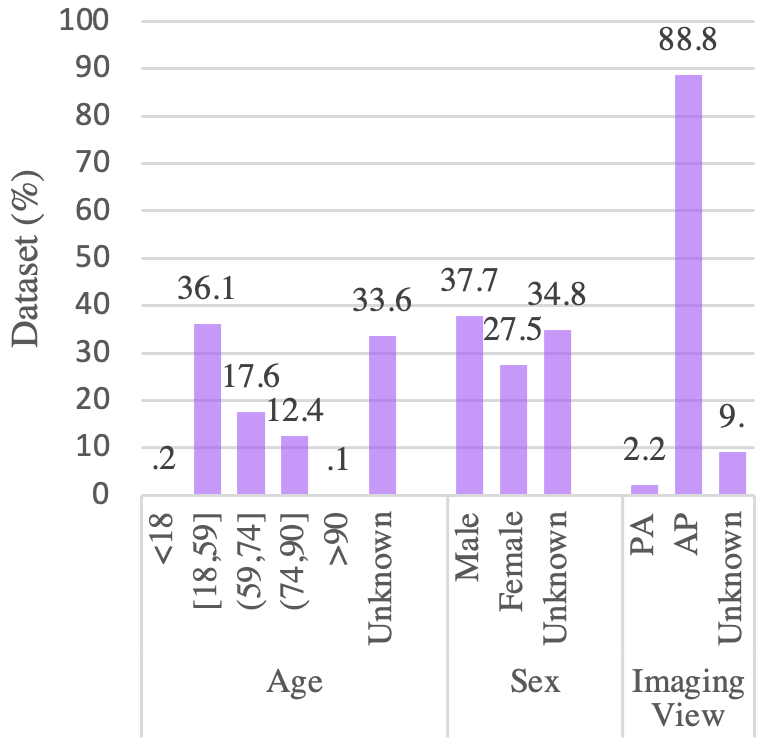}
        \caption{Distribution of demographic variables and imaging views in COVIDx~CXR-3 for SARS-CoV-2 positive patient cases and CXR images respectively.}
        \label{fig:covid_dist}
    \end{figure}

    The distribution of demographic patient data for age and sex as well as for CXR imaging view is shown across all infection types in Table~\ref{tab:demographic} and for SARS-CoV-2 positive cases only in Figure ~\ref{fig:covid_dist}. The dataset does not indicate a bias for youth or senior vulnerable age groups with the majority of patients (66.2\%) falling within the range of 18 to 59. Moreover, in terms of patients with documented sex, there is a relatively similar fraction of male (53.2\%/37.7\%) and female (40.1\%/27.5\%) patients across all infections as well as across SARS-CoV-2 positive cases respectively. In terms of imaging view, a majority of the CXR images are captured in anterior-posterior (AP) (67\%) view as opposed to posterior-anterior (PA) (28.1\%) view.

    Systems trained on the COVIDx~CXR-3 benchmark dataset have the opportunity to detect each individual documented infection class or group normal and pneumonia cases together to form a negative SARS-CoV-2 label for targeted SARS-CoV-2 screening. The hope is that the release of COVIDx~CXR-3 in an open-source manner will help encourage researchers, clinical scientists, and citizen data scientists to accelerate advancements and innovations in the fight against the pandemic.
    
    Finally, to serve as a baseline reference for comparison purposes, Table~\ref{tab:benchmarks} provides benchmark test performance for three different deep neural network architecture designs on the COVIDx~CXR-3 test dataset, as conducted in~\cite{medusa,cxr2}.  We hope that COVIDx~CXR-3 can assist scientists in advancing research against both the COVID-19 pandemic and related diseases.

    \begin{table}
        \caption{Sensitivity, positive predictive value (PPV), and accuracy of benchmark networks on the COVIDx~CXR-3 test dataset~\cite{medusa,cxr2}.}
        \label{tab:benchmarks}
        \centering
        \begin{tabular}{l l l l}
            \toprule
            \textbf{Architecture} & \textbf{Sensitivity (\%)} & \textbf{PPV (\%)} & \textbf{Accuracy (\%)} \\
            \midrule
            DenseNet201~\cite{densenet} & 82.9 & 88.9 & 90.3\\
            ResNet-50~\cite{resnetv2} & 88.5 & 92.2 & 90.5\\
            COVID-Net~\cite{covidnet} & 93.5 & \textbf{100} & 94.0 \\ 
            COVID-Net CXR-2~\cite{cxr2} & 95.5 & 97.0 &  96.3 \\
            MEDUSA~\cite{medusa} & \textbf{97.5} & 99.0 & \textbf{98.3}\\
            \bottomrule
        \end{tabular}
    \end{table}

\section*{Potential Negative Societal Impact}
    While the aim with this large-scale benchmark dataset release is to support researchers, clinicians, and citizen data scientists in advancing research, one negative societal impact that can potential arise from this release the misuse of the collected data. For example, the benchmark dataset may be utilized by insurance companies to construct machine learning algorithms for forecasting future medical expenses, and as such lead to increased insurance premiums by insurance companies that could be inappropriate for a given patient given imperfect predictive analytics.

\section*{Acknowledgements}
    We would like to thank the Canada Research Chairs program and the the Natural Sciences and Engineering Research Council of Canada (NSERC).

\bibliographystyle{unsrtnat}
\bibliography{references}

%%%%%%%%%%%%%%%%%%%%%%%%%%%%%%%%%%%%%%%%%%%%%%%%%%%%%%%%%%%%

% \appendix

% \section{Appendix}

% Optionally include extra information (complete proofs, additional experiments and plots) in the appendix.
% This section will often be part of the supplemental material.

\end{document}